\documentclass[aps,prl,preprint,superscriptaddress]{revtex4-2}

\usepackage{graphicx}
\usepackage{siunitx}
\usepackage{xcolor}
\usepackage{bm}
\usepackage{lineno}

\begin{document}


\title{Direct excitation of Kelvin waves on quantized vortices}


\author{Yosuke Minowa}
\email[]{minowa.yosuke.es@osaka-u.ac.jp}
\affiliation{Graduate School of Engineering Science, Osaka University, 1-3, Machikane-yama, Toyonaka, Osaka, Japan}

\author{Yuki Yasui}
\affiliation{Graduate School of Engineering Science, Osaka University, 1-3, Machikane-yama, Toyonaka, Osaka, Japan}

\author{Tomo Nakagawa}
\affiliation{Department of Physics, Osaka City University, 3-3-138 Sugimoto, Osaka, Japan}

\author{Sosuke Inui}
\affiliation{National High Magnetic Field Laboratory, 1800 East Paul Dirac Drive, Tallahassee, Florida 32310, USA}

\affiliation{Mechanical Engineering Department, FAMU-FSU College of Engineering, Florida State University, Tallahassee, Florida 32310, USA}

\author{Makoto Tsubota}
\affiliation{Department of Physics, Osaka Metropolitan University, 3-3-138 Sugimoto, Osaka, Japan}
\affiliation{Nambu Yoichiro Institute of Theoretical and Experimental Physics (NITEP), Osaka Metropolitan University, 3-3-138 Sugimoto, Osaka, Japan}

\author{Masaaki Ashida}
\affiliation{Graduate School of Engineering Science, Osaka University, 1-3, Machikane-yama, Toyonaka, Osaka, Japan}


\date{\today}

\begin{abstract}\textbf{
Helices and spirals, prevalent across various systems, play a crucial role in characterizing symmetry\cite{hasanColloquiumTopologicalInsulators2010}, describing dynamics\cite{funatsuImagingSingleFluorescent1995}, and imparting unique functionalities\cite{toyodaTransferLightHelicity2013,rosslerSpontaneousSkyrmionGround2006}, attributed to their inherent simplicity and chiral nature. A helical excitation\cite{thomsonVibrationsColumnarVortex1880} on a quantized vortex\cite{donnellyQuantizedVorticesHelium1991}, an example of a one-dimensional topological defect, emerges as a Nambu-Goldstone mode following spontaneous symmetry breaking\cite{kobayashiKelvinModesNambuGoldstone2014}, known as a Kelvin wave. Kelvin waves play a vital role in energy dissipation within inviscid quantum fluids\cite{barenghiQuantumTurbulence2023,vinenQuantumTurbulence2002a,vinenQuantumTurbulenceAchievements2010}. However, deliberately exciting Kelvin waves has proven to be challenging. Here, we introduce a controlled method for exciting Kelvin waves on a quantized vortex in superfluid helium-4. We used a charged nanoparticle, oscillated by a time-varying electric field, to stimulate Kelvin waves on the vortex. A major breakthrough in our research is the confirmation of the helical nature of Kelvin waves through three-dimensional image reconstruction, providing visual evidence of their complex dynamics. Additionally, we determined the dispersion relation and the phase velocity of the Kelvin wave and identified the vorticity direction, enhancing our understanding of quantum fluid behavior. This work elucidates the dynamics of Kelvin waves and pioneers a novel approach for manipulating and observing quantized vortices in three dimensions, thereby opening new avenues for exploring quantum fluidic systems.
}

\end{abstract}


\maketitle

\section*{Introduction}

In 1880, Lord Kelvin introduced the concept of helical wave excitation on a vortex line in a homogeneous, incompressible, inviscid fluid. Known as Kelvin waves, these are among the most fundamental excitations in vortices\cite{saffmanVortexDynamics1993}. Their manifestations in classical fluid dynamics have been extensively studied in various systems\cite{maxworthyWaveMotionsVortex1985a}, including those outside the laboratory, such as tornadoes\cite{dahlCentrifugalWavesTornadoLike2021}. Helical structures are a ubiquitous and intriguing research target in the wide field of science, principally owing to the fact that they represent one of the simplest and most fundamental entities representing two distinct chiralities, or handedness. Helical or spiral structures, fundamental to various phenomena, range from DNA and actin filaments\cite{funatsuImagingSingleFluorescent1995} to optical vortices\cite{toyodaTransferLightHelicity2013} and magnetic skyrmions\cite{rosslerSpontaneousSkyrmionGround2006}. The chirality defines the symmetry and imparts crucial functional properties to the system\cite{hasanColloquiumTopologicalInsulators2010}.

In superfluid, helical Kelvin waves on a quantized vortex emerge as a Nambu-Goldstone mode due to spontaneous symmetry breaking\cite{kobayashiKelvinModesNambuGoldstone2014}. Kelvin waves on quantized vortices are an optimal subject for studying vortex excitations, as their zero-viscosity nature preserves the assumptions of Kelvin's original theory, and the quantization of circulation ensures the stable existence of vortex lines. Furthermore, Kelvin waves play a crucial role in advancing our universal understanding and facilitating comparisons between classical and quantum turbulence\cite{barenghiQuantumTurbulence2023}. While at length scales larger than the mean intervortex distance in quantum turbulence, energy is transferred to smaller scales via a classical Richardson Cascade; at smaller scales, energy cascades through Kelvin waves. Here, nonlinear coupling among various Kelvin wave modes results in a net transfer of energy toward higher wave numbers\cite{kivotidesKelvinWavesCascade2001,kozikKelvinWaveCascadeDecay2004,vinenKelvinWaveCascadeVortex2003,baggaleyKelvinwaveCascadeVortex2014}. 

Despite the acknowledged importance of Kelvin waves in quantum fluid systems, research has predominantly relied on numerical simulations and theoretical analyses, with limited experimental studies\cite{bretinQuadrupoleOscillationSingleVortex2003,tellesDynamicalEvolutionDecay2021}, including observations of spontaneous Kelvin wave emission after accidental vortex reconnections\cite{fondaDirectObservationKelvin2014}. The challenge in experimental research has been the lack of a reliable method for exciting Kelvin waves in a well-controlled manner. Recently, our team demonstrated the visualization of a quantized vortex using silicon nanoparticles\cite{minowaVisualizationQuantizedVortex2022} prepared by in situ laser ablation\cite{minowaInnerStructureZnO2017}. In the current study, we pioneer a groundbreaking approach by expanding the visualization scheme to not only excite and manipulate but also observe quantized vortices in three-dimensional space for the first time. We utilized a charged silicon nanoparticle localized on a quantized vortex to excite Kelvin waves with an oscillating electric field. The helical nature of the excited Kelvin waves was confirmed via three-dimensional image reconstruction, employing a dual-camera system. Detailed analysis provided insights into the dispersion relation and phase velocity of Kelvin waves, enabling the determination of the vorticity direction.

\section*{Excitation of Kelvin waves}

\begin{figure}[h]
\includegraphics[width=1.0\columnwidth]{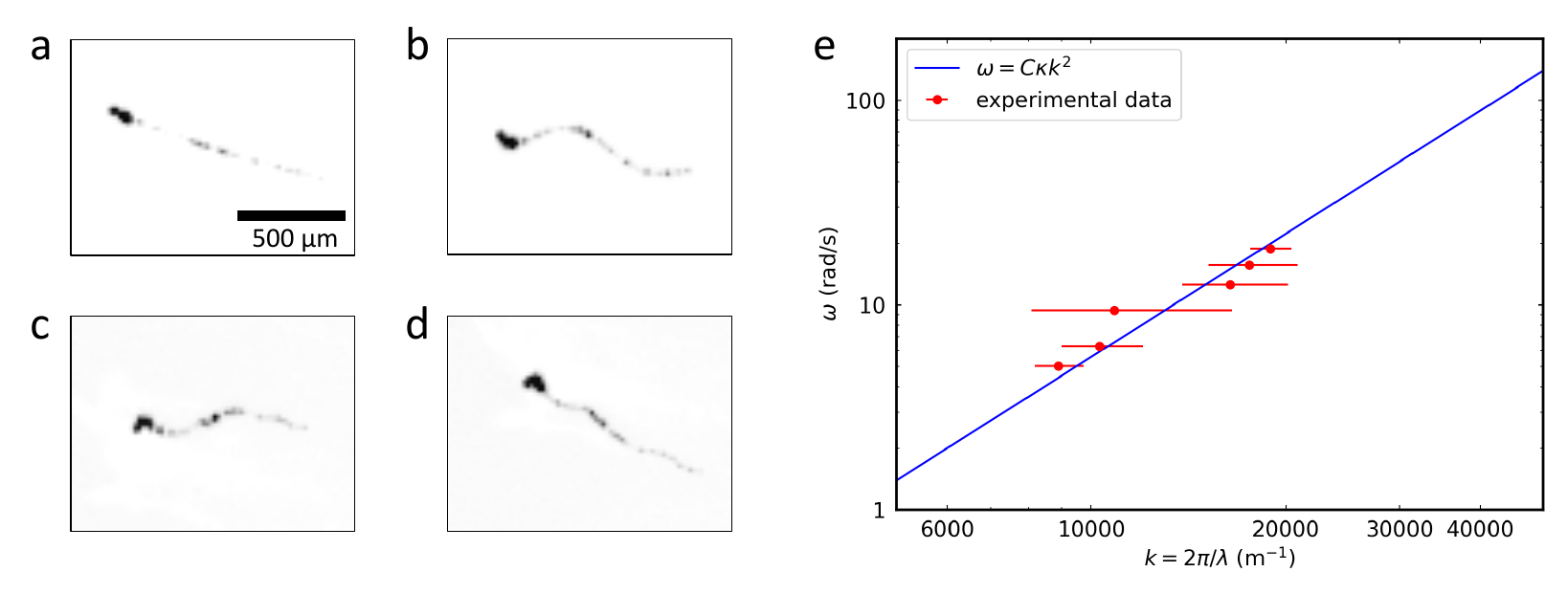}
\caption{\textbf{Kelvin wave excitation and its dispersion relation.}  a-d, Single quantized vortex visualized using silicon nanoparticles before (a) and at different stages during (b-d) the application of AC electric fields at 0.8 Hz (b), 1.5 Hz (c) and 2.5 (d) Hz. e, Dispersion relation of the excited Kelvin wave with a theoretical fit line. Each error bar indicates the 99 \% confidence interval.}
\label{fig:dispersion}
\end{figure}

 We prepared semiconductor silicon nanoparticles via in-situ laser ablation at \SI{1.4}{K} in superfluid $^4$He. Some of the produced nanoparticles are naturally attracted towards the core of quantized vortices nearby and become localized there\cite{bewleyVisualizationQuantizedVortices2006}. We observed a group of aligned silicon nanoparticles along the vortex lines, which allowed us to visualize the filament-like quantized vortex cores through light scattering, as shown in Fig. \ref{fig:dispersion}a (the image was intensity-inverted for clarity; See Methods section). The laser ablation process charges some of the produced nanoparticles\cite{taccognaNucleationGrowthNanoparticles2015}. We applied a time-varying electric field to oscillate the charged nanoparticle localized on the left end of the visualized segment of the quantized vortex, as shown in Fig. \ref{fig:dispersion}a, at $0.8$ Hz. This forced oscillation of the nanoparticle stimulated the quantized vortex to deform and emit a propagating wave along the vortex line (Fig. \ref{fig:dispersion}b). A clear sinusoidal shape was observed, and we measured the wavelength of the excitation. Repeating the excitation process with different frequencies, ranging from 0.8 Hz to 3.0 Hz (Fig. \ref{fig:dispersion}b-d), allowed us to measure the wavelength of each excitation and determine its dispersion relation, as depicted in Fig. \ref{fig:dispersion}e. The dispersion relation of Kelvin waves has been extensively studied and is well-documented in the literature\cite{brugarinoWavesVortexFilament2015,baggaleyKelvinwaveCascadeVortex2014,donnellyQuantizedVorticesHelium1991}. This relation can be summarized by the equation
 \begin{equation}
    \omega = C\kappa k^2\label{eq:dispersion},
\end{equation}
 where $\omega$ and $k$ are the angular frequency and the wavenumber of the Kelvin wave, respectively; $\kappa = h/m$ represents the circulation quanta; and $C$ is a dimensionless factor on the order of unity. The fitting of the experimental data to this equation yields $C = 0.56$ indicating that the observed dynamics of the excitation closely follow the theoretically predicted behavior of Kelvin waves. With the successful alignment of our experimental data with the theoretical prediction regarding the dispersion relation, our focus shifted towards a more detailed examination of its spatial configuration, particularly its helical nature.

 \section*{3D dynamics of Kelvin waves}
 \begin{figure}[h]
\includegraphics[width=0.95\columnwidth]{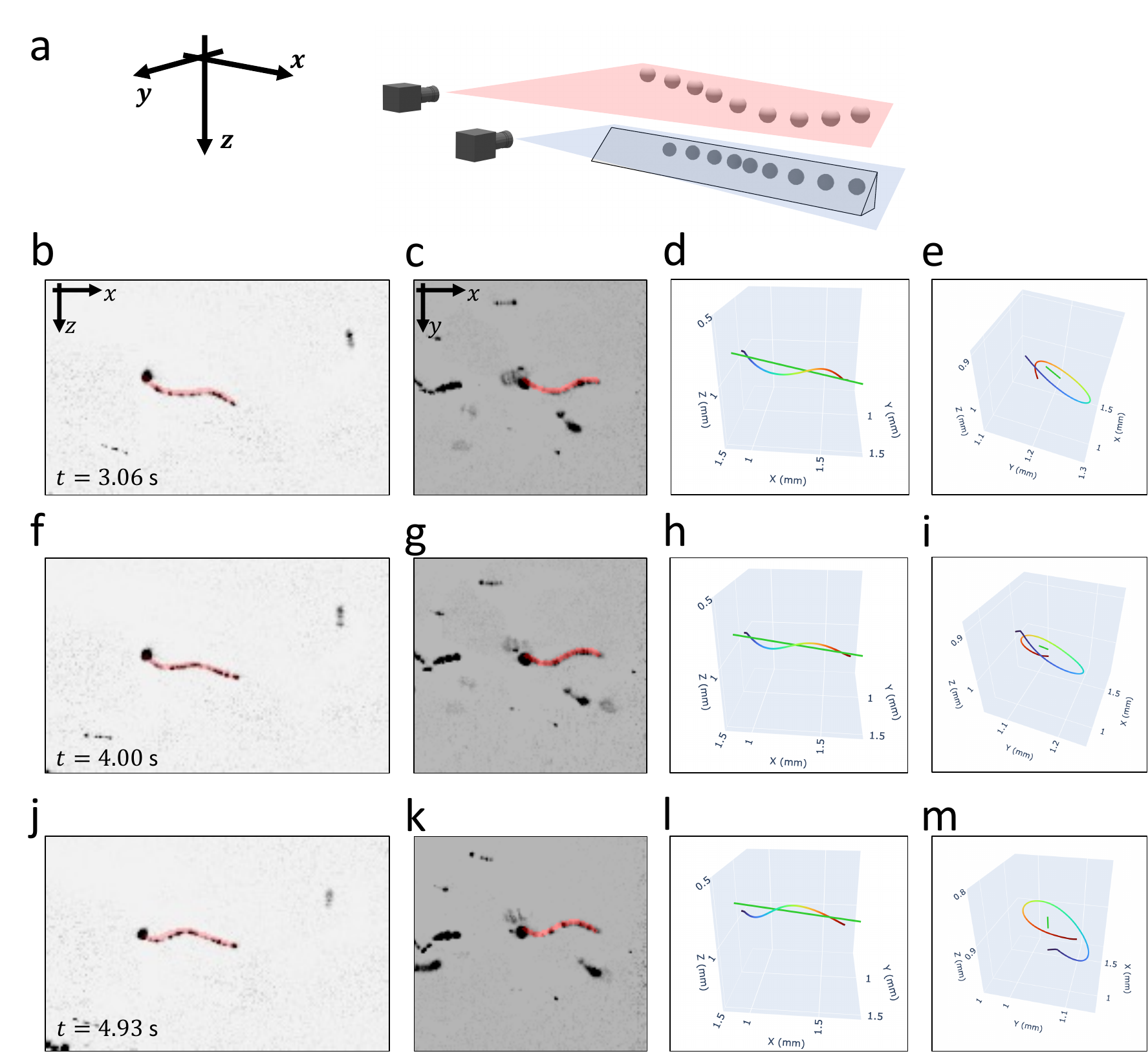}
\caption{\textbf{Experimental three-dimensional visualization of Kelvin wave dynamics along a quantized vortex} a, Schematics of the experimental setup: one camera was positioned to capture a front view of the Kelvin wave, while the second camera used a mirror to achieve a bottom view. b,f,j, Front views of the Kelvin wave at $t=3.06$ s (b), $t=4.00$ s (e), and $t=4.93$ s (h). Red curves are traced splines representing the part of the vortex line. c,g,k, Corresponding bottom views. d,e,h,i,l,m, Reconstructed three-dimensional Kelvin wave. The color of the curve varies along its length in a rainbow spectrum, corresponding to the $x$-coordinate of each point, to enhance three-dimensional recognition. Green lines are guides to clarify the handedness of the Kelvin wave (see body text). Both the front (d, h, and l) and side (e, i, and m) views are crucial for fully appreciating the helical shape: the front view more clearly reveals the handedness, while the side view distinctly showcases the circular shape.}
\label{fig:3d}
\end{figure} 
In order to examine the helical nature of the Kelvin wave, we visualized the excited wave three-dimensionally by devising an orthogonal dual-camera image reconstruction method (see Methods section for the details). We positioned a mirror at a $45^{\circ}$ angle inside the experimental cell to enable the capture of the bottom view of the quantized vortices in addition to the normal front view. We used two cameras simultaneously, one for the front and one for the bottom view, as schematically shown in Fig. \ref{fig:3d}a with the definition of the $x$, $y$ and $z$ axes. Typical snapshots of an excited wave observed from the front (Fig. \ref{fig:3d}b, f, and j) and from the bottom (Fig. \ref{fig:3d}c, g, and k) are presented. Clearly seen wavy shapes in both views indicate the helical deformation of the vortex line and imply the Kelvin wave excitation. In each snapshot, the shape of the vortex line was traced with a spline curve, which appears as red lines in Fig. \ref{fig:3d}b, c, f, g, j, and k. From the traced curve in both views, we can reconstruct the three-dimensional shape and position of the vortex line (Methods section). The corresponding reconstructed three-dimensional images of the vortex are shown in Fig. \ref{fig:3d}d, e, h, i, l, and m (see also Supplementary Video 1 and 2). The experimental three-dimensional images evidently show a helical structure, where green lines are drawn to show the center of the helix. The results confirm that the excited wave indeed corresponds to the helical Kelvin wave. Our visualization also shows that the observed Kelvin wave is left-handed. Here, we categorize the handedness of the Kelvin waves as `right-handed' or `left-handed' based on the established criteria for their static helical shapes. To further explore the origins of the handedness of the Kelvin wave, the mechanisms behind its excitation, and the effects of forced oscillation by charged nanoparticles, we next turn to numerical simulations.

\section*{Vortex filament calculations}
\begin{figure}[h]
\includegraphics[width=0.8\columnwidth]{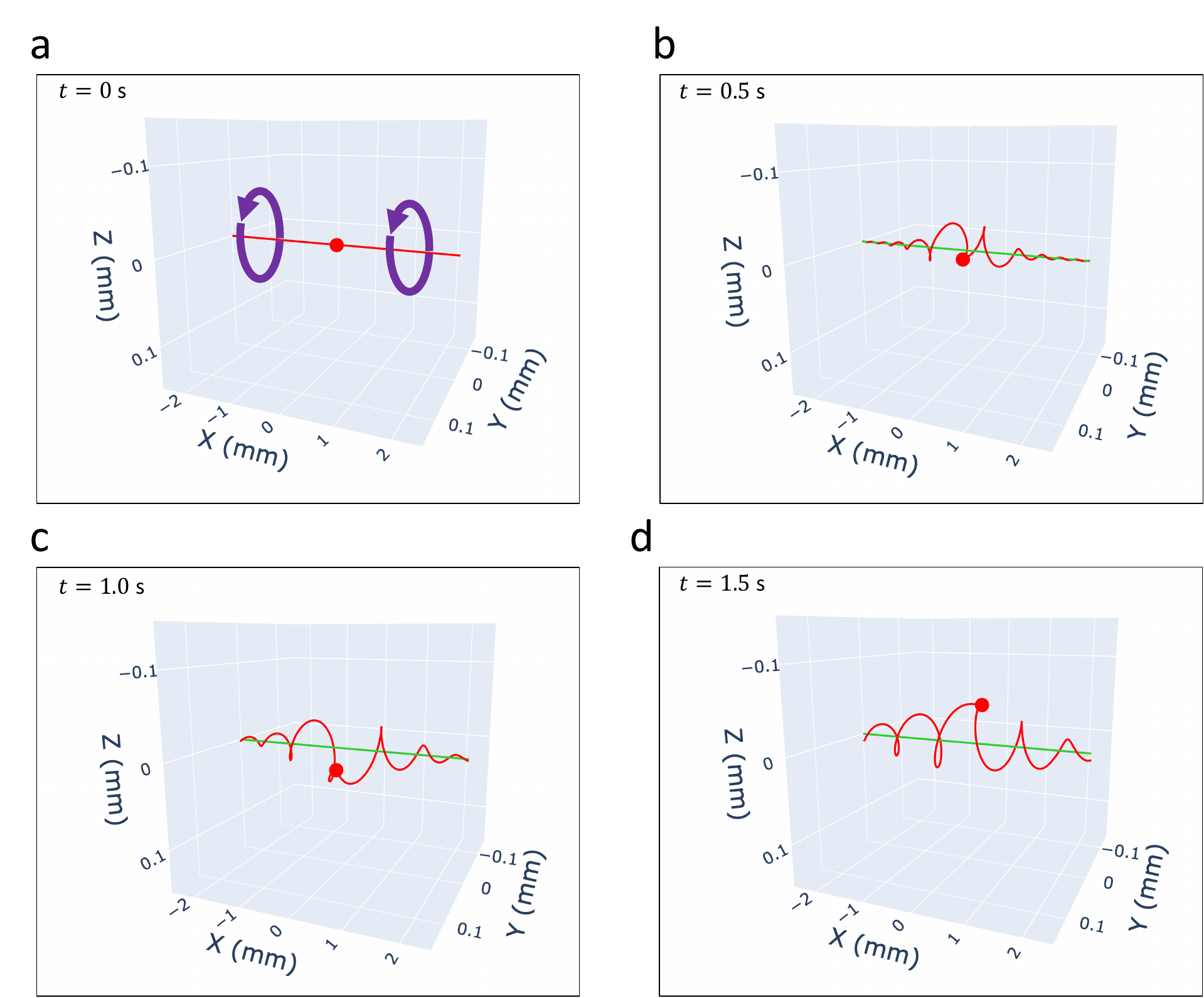}
\caption{\textbf{Numerical simulation of Kelvin wave excitation on a quantized vortex.} a, A straight quantized vortex filament before the excitation, with the red mark representing the charged nanoparticle and the purple arrow indicating the circular velocity field around the quantized vortex. b-d, Two Kelvin waves propagating in opposite directions with opposite handednesses. Green lines are provided as guides to clarify the handedness of the Kelvin wave.}
\label{fig:simulation}
\end{figure}

To model the excitation of Kelvin waves through the forced oscillation of a charged nanoparticle, we conducted a numerical simulation involving a vortex filament. The vortex filament model (VFM) serves as an effective framework for examining the dynamics of quantized vortices in superfluid $^4$He\cite{donnellyQuantizedVorticesHelium1991,schwarzThreedimensionalVortexDynamics1985,tsubotaNumericalStudiesQuantum2017}. This model simplifies the extremely thin vortex core ($\sim 1\ \mathrm{\AA}$) to a one-dimensional filament, disregarding the vortex core's intricate structure. The VFM is particularly efficient when the vortex core's size is negligible compared to the system's typical dimensions (in our case $\sim 1\ \mathrm{cm}$), which allowed us to model the experimental conditions and elucidate the dynamics by solving the equations of motion for both the vortex filament and the charged particle confined to the vortex. The initial setup featured two parallel solid boundaries perpendicular to the $x$ axis, spaced $1\ \mathrm{cm}$ apart ($x=\pm 5$ mm), with a straight line vortex aligned in the positive $x$-direction and normal to these boundaries, as illustrated in Fig. \ref{fig:simulation}a. The temperature in the simulations was maintained at 1.4 K, with the charged particle trapped at the vortex's midpoint.  To simulate the experimental conditions, we applied an AC electric field in the $z$-direction with an amplitude of $8823\ \mathrm{V/m}$ and a frequency of $0.8\ \mathrm{Hz}$. We also used the typical particle size of $87\ \mathrm{nm}$\cite{minowaVisualizationQuantizedVortex2022}. Since the charges of particles produced by laser ablation process are generally non-uniform and strongly dependent on experimental parameters, such as laser light pulse energy\cite{dellaglioInvestigationMaterialPlasma2019}, and are difficult to measure in situ, we assumed a charge of $1000e=-1.602\times 10^{-16}\ \mathrm{C}$ for the simulation. This value was chosen to align the simulated amplitude of the excited wave with the experimental observations.

The red mark in Fig. \ref{fig:simulation} represents the charged nanoparticle localized on the vortex filament, depicted as a red line. From the initial straight vortex line (Fig. \ref{fig:simulation}a), the nanoparticle subjected to the AC electric field induces the emission of helical Kelvin waves in both directions simultaneously. The simulation results confirm the excitation of Kelvin waves by a charged nanoparticle and closely replicate the experimental observations. It is important to note that in our experimental results, shown in  Fig. \ref{fig:dispersion} and \ref{fig:3d}, the charged nanoparticles under the electric field were positioned at one end of the visualized segment of the quantized vortex, leading to the observation of a single Kelvin wave emission in each case. In the simulated results, the two emitted Kelvin waves exhibit opposite handedness: the right-handed Kelvin wave propagates in the positive $x$-direction, while the left-handed one propagates in the negative $x$-direction. The handedness and propagation direction of the Kelvin wave, along with its vorticity direction (indicated by purple arrows in Fig. \ref{fig:simulation}a), are interrelated. Since we can determine the handedness and propagation direction of a Kelvin wave in experimental observations, we are able to identify the vorticity direction. Having already established the handedness of the Kelvin wave as left-handed in Fig. \ref{fig:3d}, we now focus on a detailed examination of its propagation.

\section*{Phase velocity and vorticity direction}
\begin{figure}[h]
\includegraphics[width=1.0\columnwidth]{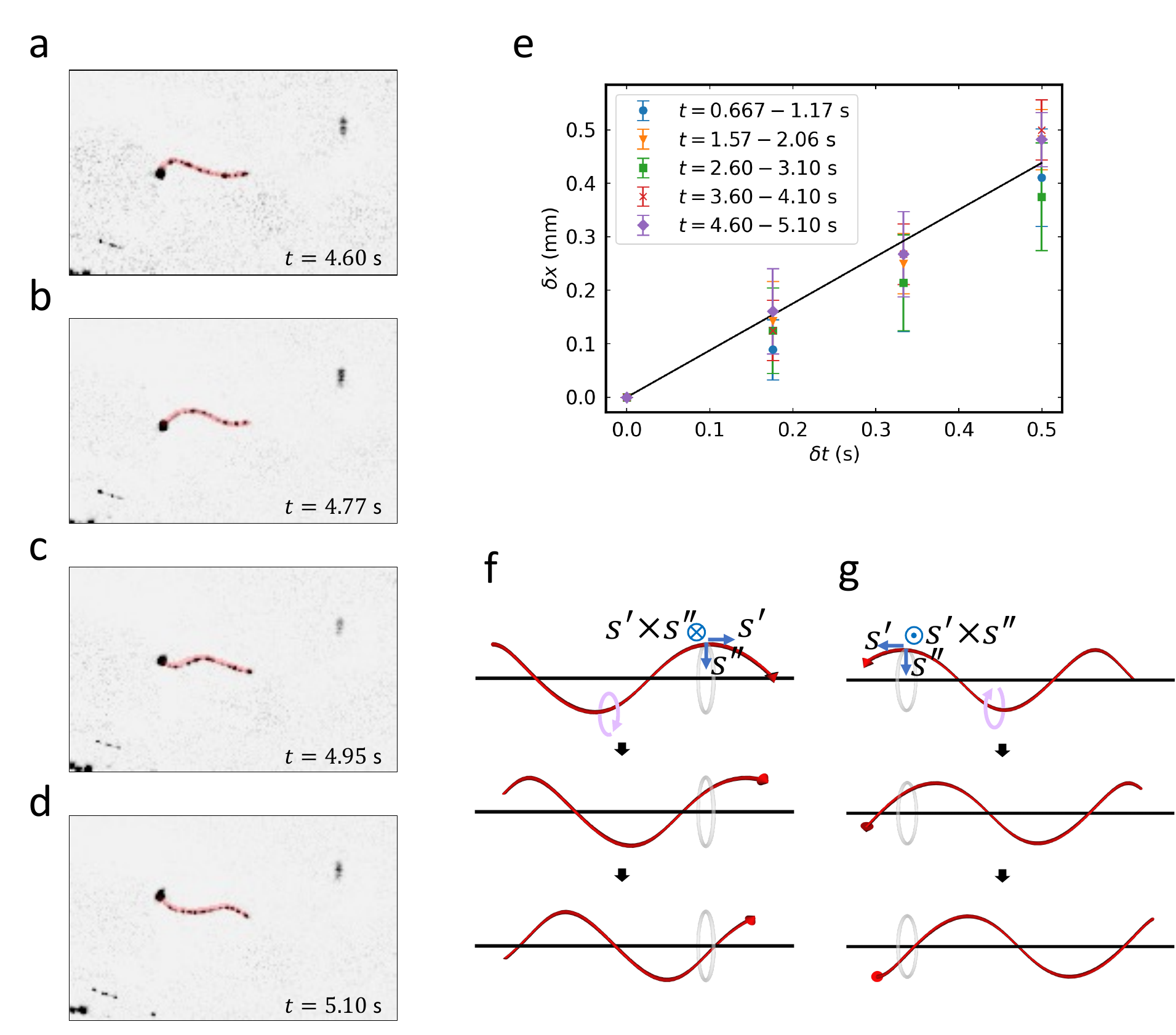}
\caption{\textbf{Propagation of Kelvin waves.} a-d, the Kelvin wave shown in Fig. \ref{fig:3d} propagating to the right. e, Position of the phase front as a function of time. f, g. Propagation of right-handed (f) and left-handed (g) Kelvin waves. Red arrowheads and purple arrows  indicate the directions of the vorticity. A right-handed Kelvin wave propagates parallel to the vorticity direction(f), while a left-handed Kelvin wave propagates anti-parallel to the vorticity direction(g).}
\label{fig:velocity}
\end{figure}

Figure \ref{fig:velocity}a-d presents a sequence of images showcasing the dynamics of the quantized vortex observed in Fig. \ref{fig:3d}, clearly demonstrating the rightward propagation of the Kelvin wave. The propagation speed, or phase velocity, of the Kelvin wave was deduced from the dynamics of the phase front position and is depicted in Fig. \ref{fig:velocity}e. The movement of the phase front, over a duration of 500 ms, is plotted at intervals of every 167 ms. From the linear fitting, the phase velocity of the Kelvin wave is determined to be 0.9 mm/s. A phase velocity of a wave can be deduced from its dispersion relation. Starting from the dispersion relation equation (\ref{eq:dispersion}), we can calculate the theoretically expected phase velocity $v_p$ as the slope of the dispersion
\begin{equation}
    v_p=\frac{\omega}{k} = C\kappa k\label{eq:velocity}.
\end{equation}
As in our experiment $k\sim$ $10^4$ m$^{-1}$, we expect $v_p\sim 1$ mm/s which is remarkably consistent with the experimental result. Now, we examine the relationship between the propagation direction, the handedness of a Kelvin wave, and the vortcity direction. 
When we have a vortex line, there are two possible directions of vorticity. For each vorticity, Kelvin waves on the vortex can be either right-handed or left-handed. Among the four possible Kelvin waves, only two propagate to the right, as schematically shown in Fig \ref{fig:velocity}f-g. The directions of vorticity are illustrated as red arrowheads and purple arrows, which determine the direction of the tangent vector $\bm{s}'$, where $\bm{s}$ refers to a point on the vortex line\cite{schwarzThreedimensionalVortexDynamics1985} and the prime denotes differentiation with respect to the length of the vortex line. The curvature of the vortex line is indicated by the principal normal vector $\bm{s}''$. The vectors $\bm{s}'$ and $\bm{s}''$ define the handedness of the Kelvin waves, as well as the self-induced velocity, which is proportional to $\bm{s}'\times\bm{s}''$. This self-induced velocity, in turn, drives the circular motion of an infinitesimal segment of the vortex line, as illustrated by the gray orbits in Fig \ref{fig:velocity}f and g. This mechanism determines the propagation direction of the Kelvin waves to be to the right, as demonstrated in the cases of Fig \ref{fig:velocity}f and g. Given that the Kelvin wave shown in Fig \ref{fig:3d} is left-handed and propagates to the right, this experimental observation confirms that the vorticity direction of the quantized vortex observed matches that shown in Fig \ref{fig:velocity}g.

\section*{Discussion and outlook}
Kelvin waves have been a central topic in the field of quantum fluid dynamics, yet there has been no direct method to excite them until now. Our study would fundamentally transform the landscape of quantum fluid research by introducing a method to excite these waves, along with a three-dimensional visualization technique to reveal the detailed dynamics of quantized vortices. Furthermore, we have determined the direction of vorticity based on experimental observations, representing the first such demonstration outside of specialized scenarios, such as those involving rotating cryostats\cite{makinenRotatingQuantumWave2023} that generate aligned quantized vortices with predetermined vorticity.

Kelvin waves are ubiquitous and have been studied in numerous systems, including cyclones\cite{dahlCentrifugalWavesTornadoLike2021}, neutron stars\cite{linkVortexPinningNeutron2022}, plasma\cite{nakariakovKinkOscillationsCoronal2021}, and superconductors\cite{degennesCollectiveModesVortex1964}. They represent one of the most fundamental excitations and characterize the elemental properties of vortices or one-dimensional phase singularities. Consequently, the physics of vortices often dominates the dynamics of systems containing them\cite{nakamuraNonreciprocalTerahertzSecondHarmonic2020}. However, detailed experimental study of Kelvin waves on vortices has seen limited advancement, largely due to the lack of a reliable protocol for their excitation and observation. Therefore, our study could significantly impact the investigation of any system featuring vortices or one-dimensional phase singularities\cite{kinjoSuperconductingSpinSmecticity2022}.

Finally, our approach, which utilizes charged nanoparticles on a quantized vortex core, opens up new avenues not only for exciting quantized vortices but also for manipulating and controlling their motion\cite{minowaOpticalTrappingNanoparticles2022}. The precise manipulation and control of vortices are essential for studying the interactions among multiple quantized vortices\cite{minowaVisualizationQuantizedVortex2022}. The remote and instantaneous nature of electric-field-driven manipulation enables us to uncover fundamental properties of quantum fluid dynamics\cite{tangSuperdiffusionQuantizedVortices2021,skrbekPhenomenologyQuantumTurbulence2021,mantiaQuantumClassicalTurbulence2014}.

\section*{methods}

\subsection*{Silicon nanoparticles and quantised vortex visualization in superfluid $^4$He}
We placed a single crystal silicon wafer (single side polished) in the experimental cell. The cell was a 3 cm $\times$ 3 cm $\times$ 3 cm acrylic box, three surfaces of which were equipped with fused silica windows for laser introduction. Focused 6-ns laser light pulses with a wavelength of 355 nm were irradiated onto the wafer at a repetition rate of 10 Hz for 15 s. The laser ablation process resulted in the production and dispersion of many silicon nanoparticles within the cell, the dynamics of which could be observed through the scattering of illuminated laser light with a wavelength of 532 nm and a power of 500 mW. After the laser ablation, we observed numerous samples of aligned silicon nanoparticles, which visualized quantized vortices. The recorded images were subjected to background subtraction and intensity inversion for further analysis. 

\subsection*{Forced oscillation of a charged nanoparticle and three dimensional image reconstruction}
To apply an alternating current (AC) electric field, we positioned a pair of transparent electrodes—indium tin oxide films on glass substrates—in a horizontal arrangement above and below the observation volume, which contained numerous quantized vortices. An AC electric voltage of ±150 V was applied between the electrodes once the flow, initiated by laser ablation, had stabilized and after identifying long quantized vortices (approximately 1 mm in length); typically, a waiting period of 100 seconds after the laser ablation was observed before applying the electric voltage. A mirror, measuring 15 mm by 5 mm, was positioned directly beneath the electrodes at a 45-degree angle to facilitate a bottom view of the quantized vortices. Two cameras were used to capture the experiment: one recorded the front view, and the other recorded the bottom view, both at a frame rate of 30.029 fps.

The motion of the quantized vortex during excitation was traced separately for each view. The shape of the vortex line was delineated using a spline curve to interpolate between discrete nanoparticles. By combining the $z$-coordinates from the front view and the $y$-coordinates from the bottom view, we were able to reconstruct the three-dimensional appearance of the quantized vortex for each frame.

\subsection*{Vortex filament calculations}
The vortex filament model simplifies vortex motion by reducing the detailed structure of the vortex core to one-dimensional filaments. To compute their motion, each filament is discretized into a series of connected points. At $0\ \mathrm{K}$, the force acting on the filament comprises the vortex tension $\bm{F}_T$ and the Magnus force $\bm{F}_M$. At finite temperature, superfluid $^4$He is described by the two-fluid model, consisting of inviscid superfluid component and viscous normal fluid component arising from thermal excitations\cite{tiszaTransportPhenomenaHelium1938,landauTheorySuperfuidityHelium1941}. In this context, the drag force $\bm{F}_D$ acts between vortex cores and normal fluid. Denoting a position of the filament by $\bm{s}(\xi)$, parameterized with its arclength $\xi$, the force $\bm{F}_T$, $\bm{F}_M$ and $\bm{F}_D$ acting at that position are
\begin{eqnarray}
\bm{F}_T&=&\rho_s\kappa\beta_\mathrm{ind}[\bm{s}^\prime(\xi+\Delta\xi/2)-\bm{s}^\prime(\xi-\Delta\xi/2)]\label{eq:tension}\\
\bm{F}_M&=&\rho_s\kappa\bm{s}^\prime\times(\bm{v}_L-\bm{v}_{s,nl})\Delta\xi\label{eq:Mineda_magnus}\\
\bm{F}_D&=&\left[\gamma_0\bm{s}^\prime\times\bm{s}^\prime\times(\bm{v}_L-\bm{v}_n)-\gamma_0^\prime\bm{s}^\prime\times(\bm{v}_L-\bm{v}_n)\right]\Delta\xi,
\end{eqnarray}
where $\rho_s$, $\bm{v}_L$, $\bm{v}_n$ and $\bm{v}_{s,nl}$ are the superfluid density, the velocity of the filament at $\bm{s}(\xi)$, the normal-fluid velocity and the non-local velocity obtained by computing the Biot-Savart integral along the filament over its arclength $\xi_1$ except for the neighborhood of $\xi_1 = \xi$. $\Delta \xi$ is the distance from adjacent point, and  $\bm{s}^\prime$ refers to $\frac{d\bm{s}}{d\xi}$, which is a tangent vector at $\bm{s}(\xi)$. $\beta_\mathrm{ind}=\frac{\kappa}{4\pi}\ln\left[\frac{2\sqrt{l_+l_-}}{\sqrt{e}a}\right]$ is determined by the geometric mean $\sqrt{l_+l_-}$ of the distances from $\bm{s}(\xi)$ to both neighboring points and the vortex core radius $a$. $\gamma_0$ and $\gamma_0^\prime$ are temperature-dependent drag coefficients\cite{donnellyQuantizedVorticesHelium1991}. 

Assuming the effective mass $m_\mathrm{eff}$ of the vortices is sufficiently small, we can ignore the inertial term in the equation of motion for the filaments $m_\mathrm{eff} \frac{d^2\bm{s}}{dt^2}=\bm{F}_T+\bm{F}_M+\bm{F}_D$. Consequently, the equation of motion is simplified;

\begin{eqnarray}\label{eq:veom}
	\frac{d\bm{s}}{dt}&=&\bm{v}_L=\bm{v}_{s0}+\alpha\bm{s}^\prime\times(\bm{v}_n-\bm{v}_{s0})-\alpha^\prime\bm{s}^\prime\times\left[\bm{s}^\prime\times(\bm{v}_n-\bm{v}_{s0})\right]\\
	\bm{v}_{s0}&=&\frac{\kappa}{4\pi}\int_\mathcal{L}\frac{\bm{s}^\prime(\xi_1)\times(\bm{s}(\xi_1)-\bm{s}(\xi))}{|\bm{s}(\xi_1)-\bm{s}(\xi)|^3}d\xi_1,
\end{eqnarray}
where $\alpha$ and $\alpha^\prime$ are coefficients determined by $\gamma_0$, $\gamma_0^\prime$, $\rho_s$ and $\kappa$; the integral is along the all vortex filaments $\mathcal{L}$. In addition, a charged particle trapped on the vortex is subject to the Stokes drag $\bm{F}_S$ and electric force $\bm{F}_E$;
\begin{equation}
	\bm{F}_S=6\pi R\eta(\bm{v}_n-\bm{v}_p),\quad \bm{F}_E=q\bm{E},
\end{equation}
where the $R$, $\eta$ and $q$ are the radius of particle, viscosity of normal fluid and the charge of particle, respectively. $\bm{E}$ is the applied electric field. Thus, letting $M$ denote the particle's mass, an equation of motion for the particle at position $\bm{x}_p$ is given by $M\frac{d^2\bm{x}_p}{dt^2}=\bm{F}_T+\bm{F}_M+\bm{F}_D+\bm{F}_S+\bm{F}_E$, where $\bm{v}_L$ in $\bm{F}_M$ and $\bm{F}_D$ becomes the particle velocity $\bm{v}_p$\cite{minedaVelocityDistributionsTracer2013,tangImagingQuantizedVortex2023}. In this simulation, the inertial term $M\frac{d^2\bm{x}_p}{dt^2}$ can be neglected due to the small mass\cite{minowaVisualizationQuantizedVortex2022}. Finally, we solve for $\bm{v}_p$ by the equation $0=\bm{F}_T+\bm{F}_M+\bm{F}_D+\bm{F}_S+\bm{F}_E$, and then integrate the following simple differential equation temporally
	\begin{equation}\label{eq:peom}
	\frac{d\bm{x}_p}{dt}=\bm{v}_p.
	\end{equation}
The separation distance between two neighboring discretized points on a vortex filament is set between $\Delta\xi_\mathrm{min}=2\ \si{\micro m}$ and $\Delta\xi_\mathrm{max}=5\ \si{\micro m}$. The points are removed or added to satisfy the separation thresholds. The differential equations(Eq. (\ref{eq:veom}) and (\ref{eq:peom})) are solved using a fourth-order Runge-Kutta method for the temporal evolution, and the temporal resolution $\Delta t$ is $6.25\ \si{\micro s}$.

\section*{Data availability}
The data that support the findings of this study are available from the corresponding author upon reasonable request.

\section*{Acknowledgements}
This work was supported by the MEXT/JSPS KAKENHI grant numbers JP22H05139, JP23K03282, JP23KJ1832 and JP23K03305, and by JST PRESTO grant number JPMJPR1909, Japan.

\section*{Author contributions}
\subsection*{Contributions}Y.M. conceived, designed, and guided the whole project. Y.M. and Y.Y. performed the experiments and analysed the data. T.N., S.I., and M.T. conducted the vortex filament simulation and the analysis. M.A. provided technical support. Y.M. wrote the paper with inputs from all co-authors.
\subsection*{Corresponding author}
Correspondence and requests for materials should be addressed to Yosuke Minowa.

\section*{Ethics declarations}
\subsection*{Competing interests}
The authors declare no competing interests.

\section*{Additional Information}
Supplementary Information is available for this paper.

\bibliographystyle{naturemag}
\bibliography{kelvinwave}

\end{document}